\documentclass[
  a4paper,
  twocolumn,
  longbibliography,
  aps,
  pra,
  superscriptaddress,
  amsfonts,amsmath,amssymb,
  floatfix
]{revtex4-2}

\usepackage[
  colorlinks,
  citecolor=blue,
  linkcolor=blue,
  urlcolor=blue
]{hyperref}
\usepackage{graphicx,epstopdf,bm}

\usepackage{braket}
\usepackage{blindtext}
\usepackage{enumitem}
\usepackage[capitalize]{cleveref}
\usepackage{xcolor}
\usepackage[all]{hypcap}

\crefname{section}{Sec.}{Secs.}
\Crefname{section}{Section}{Sections}
\crefname{appendix}{App.}{Appces.}
\Crefname{appendix}{Appendix}{Appendices}

\DeclareMathOperator{\Tr}{Tr}
\newcommand{\e}[1]{\mathrm{e}^{#1}}
\newcommand{\op}[1]{\hat{#1}}
\newcommand{\opdag}[1]{\hat{#1}^\dagger}
\newcommand{\supop}[1]{\mathcal{#1}}
\newcommand{\trace}[2][]{\Tr_{#1}\left[#2\right]}
\newcommand{\del}[2][]{{\frac{\partial #1}{\partial #2}}}

\newcommand{\deriv}[2][]{{\frac{\mathrm{d} #1}{\mathrm{d} #2}}}

\newcommand{\difd}{\mathrm{d}}

\newcommand{\expval}[1]{{\langle#1\rangle}}
\newcommand{\Expval}[1]{{\left\langle#1\right\rangle}}

\newcommand{\cf}{cf.\ }
\newcommand{\eg}{e.g., }
\newcommand{\ie}{i.e., }

\newcommand{\figref}[2][]{Fig.~\hyperref[#2]{\ref*{#2}#1}}
\newcommand{\figureref}[2][]{Figure~\hyperref[#2]{\ref*{#2}#1}}

\begin{document}

\title{Quantum Optimal Control of Squeezing in Cavity Optomechanics
}


\author{Anton Halaski}
\affiliation{Dahlem Center for Complex Quantum Systems and Fachbereich Physik,
  Freie Universit\"{a}t Berlin, Arnimallee 14, D-14195 Berlin, Germany
}

\author{Matthias G. Krauss}
\affiliation{Dahlem Center for Complex Quantum Systems and Fachbereich Physik,
  Freie Universit\"{a}t Berlin, Arnimallee 14, D-14195 Berlin, Germany
}

\author{Daniel Basilewitsch}
\affiliation{Dahlem Center for Complex Quantum Systems and Fachbereich Physik,
  Freie Universit\"{a}t Berlin, Arnimallee 14, D-14195 Berlin, Germany
}

\author{Christiane P. Koch}
\email{christiane.koch@fu-berlin.de}
\affiliation{Dahlem Center for Complex Quantum Systems and Fachbereich Physik,
  Freie Universit\"{a}t Berlin, Arnimallee 14, D-14195 Berlin, Germany
}


\begin{abstract}
  Squeezing is a non-classical feature of quantum states that is a useful resource, for example in quantum sensing of mechanical forces. Here, we show how to use optimal control theory to maximize squeezing in an optomechanical setup with two external drives and determine how fast the mechanical mode  can be squeezed. For the autonomous drives considered here, we find the inverse cavity decay to lower-bound the protocol duration. At and above this limit, we identify a family of protocols leveraging a two-stage control strategy, where the mechanical mode is cooled before it is squeezed. 
  Identification of the control strategy allows for two important insights -- to determine the factors that limit squeezing and to simplify the time-dependence of the external drives, making our protocol readily applicable in experiments.
\end{abstract}


\maketitle

\section{Introduction}
  \label{sec:introduction}

 Squeezing refers to reducing the uncertainty in one observable at the expense of increasing the uncertainty in a different, noncommuting observable~\cite{Walls08}. Squeezed states are an important resource for quantum-enhanced sensing~\cite{CavesPRD1981, LawrieAP19} and metrology~\cite{PezzeRMP2018}. Recent applications include dark matter searches~\cite{MalnouPRX2019,BackesNature2021}, detecting motional displacement and electric fields of trapped ions~\cite{GilmoreScience2021}, improving quantum nondemolition readout~\cite{QinPRL2022}, and amplification of interactions~\cite{LiPRL2020,GroszkowskiPRL2020,VilliersPRXQ2024,BurdPRXQ2024}.

Generation of squeezing requires nonlinearity~\cite{Walls08}. It can be engineered by nonlinear drives or via coupling to another, typically driven, quantum system. A popular platform that realizes the latter paradigm is cavity optomechanics where a mechanical oscillator is coupled to an optical or microwave cavity~\cite{AspelmeyerRMP14,Clerk20}. A drive on the optical or microwave cavity can be used to engineer essentially arbitrary quantum states of the mechanical oscillator~\cite{AspelmeyerRMP14}, including strongly squeezed states. Remarkably, already at finite temperature, nonclassical states of the mechanical oscillator display useful quantum properties.
Squeezing can be realized via a multitude of protocols, leveraging both unitary~\cite{ArenzQuantum2020} and dissipative dynamics, for example using two-photon drives~\cite{KronwaldPRA13,WollmanS15}. In the latter case, the squeezed state is approached as the steady state. Dissipative protocols come with the advantage that the mechanical oscillator gets cooled and squeezed at the same time. On the downside, the approach to the steady state is typically slow.

A very fast way to achieve squeezing in a harmonic oscillator is a sudden change of the oscillator frequency which projects the ground state into a squeezed state~\cite{XinPRL2021}. While successful with trapped atoms, this protocol is hampered in cavity optomechanics by the fact that one can neither easily prepare the mechanical oscillator in its ground state nor quickly change its frequency. On the other hand, the dissipative preparation of squeezed states via two-photon driving~\cite{KronwaldPRA13,WollmanS15} can be made faster by modulating the drive amplitudes~\cite{BasilewitschAQT19}. The time-dependent shapes of the modulation were derived with optimal control theory~\cite{GlaserEPJD15,KochEPJQT22}, targeting a given (mixed) squeezed state. However, it is typically not the exact squeezed state that matters most in applications such as force sensing, but rather the amount of squeezing that can be realized.

Here, we ask, using optimal control theory~\cite{GlaserEPJD15,KochEPJQT22}, what is the maximum squeezing that can be achieved in the mechanical oscillator when optimizing the shapes of the dissipative two-photon protocol. To this end, we change perspective compared to Ref.~\cite{BasilewitschAQT19}, targeting the minimal variance of one of the mechanical oscillator's quadratures rather than a specific (squeezed) state. An additional particular benefit of optimal control theory is that it allows one to determine the minimal duration for the successful realization of a certain task, expressed by the optimization functional~\cite{CanevaPRL09,GoerzJPB11}. This bound is also referred to as the quantum speed limit \cite{BhattacharyyaJPMG83, MargolusPNP98, LevitinPRL09}. Knowing its value is of great interest in quantum information tasks where maximizing resource efficiency is crucial~\cite{Deffner_2017}. In cases where it is not possible to derive the quantum speed limit analytically, it can be determined numerically~\cite{CanevaPRL09,GoerzJPB11,dasilva2024gate}.  Here, we adopt the approach of Refs.~\cite{CanevaPRL09,GoerzJPB11} and conduct optimizations for various protocol durations, assuming a specific threshold for the target functional below which the task is considered to be fulfilled. The smallest final time in which this threshold can still be met then represents an estimate for the quantum speed limit~\cite{CanevaPRL09,GoerzJPB11}.

The remainder of the paper is organized as follows. Section~\ref{sec:Model} briefly reviews the model for the optomechanical system and two-drive protocol~\cite{KronwaldPRA13} and summarizes the essentials of optimal control theory, in particular when applied to maximize squeezing. Our optimization results are presented in Sec.~\ref{sec:results}, analyzing them in terms of the control strategies and protocol duration, respectively quantum speed limit. Section~\ref{sec:conclusion} concludes.


\section{Theoretical Framework}
  \label{sec:Model}

  \subsection{Model}

    We examine an optomechanical system \cite{LawPRA95} in which an optical cavity of frequency $\omega_\mathrm{cav}$ is coupled to a mechanical resonator with frequency $\Omega$. The Hamiltonian including a time-dependent drive $\op{H}_\text{dr}(t)$ reads
    \begin{equation}
        \op{H}(t)
        = \hbar \omega_\text{cav} \op{d}^\dagger \op{d} 
        + \hbar \Omega \op{b}^\dagger \op{b}
        - \hbar g_0 \op{d}^\dagger \op{d} (\op{b}^\dagger + \op{b}) + \op{H}_\text{dr}(t).\label{eq:H_with_H_drive}
    \end{equation}
    Here, $\op{d}$ and $\op{b}$ are the respective annihilation operators for the cavity and resonator and $g_0$ represents the optomechanical coupling strength. 
    As shown in Ref.~\cite{KronwaldPRA13}, achieving highly squeezed states is possible by driving both mechanical sidebands at frequencies $\omega_\pm = \omega_\mathrm{cav} \pm \Omega$. We therefore consider the following time-dependent drive,
    \begin{equation}
        \op{H}_\text{dr}(t)
        = \hbar \left(\alpha_+(t) e^{-i\omega_+ t} + \alpha_-(t) e^{-i\omega_- t}\right) \op{d}^\dagger + \text{H.c.},
        	\label{eq:H_dr}
    \end{equation}
    where $\alpha_\pm(t) = \pm\Omega \Bar{a}_\pm(t)$ and $\Bar{a}_\pm(t)$ denotes the field amplitude of the coherent light field. In contrast to the protocol presented in Ref.~\cite{KronwaldPRA13} and similar to Ref.~\cite{BasilewitschAQT19}, we take the amplitudes $\Bar{a}_\pm(t)$ to be time-dependent.

    To simplify the problem, we invoke the same transformations as presented in Ref.~\cite{KronwaldPRA13}, neglecting extra driving terms, which only affect classical expectation values, but not the squeezing, see \Cref{appendix:derivation} for details. The approximated Hamiltonian reads
    \begin{alignat}{2}
        \op{H}(t) 
         = &- \hbar \op{d}^\dagger \left[G_+(t) \op{b}^\dagger + G_-(t) \op{b} \right] + \text{H.c.} \nonumber\\
        &- \hbar \op{d}^\dagger \left[G_+(t) \op{b} e^{-2i\Omega t} + G_-(t) \op{b}^\dagger e^{2i\Omega t}\right] & &+ \text{H.c.},
        	\label{eq:H_final_with_rotating_terms}
    \end{alignat}
    where we have introduced rescaled pulse amplitudes $G_\pm(t) = g_0 \Bar{a}_\pm(t)$.

    We explicitly account for dissipation in both resonator and cavity. The temperature of the bath to which the cavity couples is effectively zero while the resonator is immersed in a thermal bath with thermal occupancy $n_\mathrm{th}$. The time evolution of the system in the Markov approximation is described by a master equation \cite{Breuer02},
    \begin{align}
        \deriv{t} \op{\rho}(t)
        &= -\frac{i}{\hbar}[\op{H}(t), \op{\rho}(t)] \nonumber\\
        & \qquad + \sum_{l=1}^{3} 
        \left(\op{L}_l \op{\rho}(t) \op{L}_l^\dagger - \frac{1}{2} \left\{\op{L}_l^\dagger \op{L}_l, \op{\rho}(t)\right\} \right)\label{eq:Lindblad-Master}\\
        &= \supop{L}\op{\rho}(t)\nonumber
	\end{align}
    with Lindblad operators $\op{L}_1=\sqrt{\kappa} \op{d}$, $\op{L}_2 = \sqrt{\Gamma n_\mathrm{th}} \op{b}^\dagger$ and $\op{L}_3 = \sqrt{\Gamma (n_\mathrm{th} +1)} \op{b}$. Here, $\kappa$ and $\Gamma$ are the decay rates for cavity and resonator, respectively, and $\supop{L}$ is the Liouvillian superoperator.


  \subsection{Generation of squeezed states}
    \label{subsec:theory_squeezing-generation}
    
    In order to understand the control strategies presented below, we review how squeezing is generated, in particular the roles of the two amplitudes $G_+$ and $G_-$, following Ref.~\cite{KronwaldPRA13}.
    We first consider the Hamiltonian in \cref{eq:H_final_with_rotating_terms} in the rotating wave approximation (RWA), \ie without the terms rotating at frequency $2\Omega$,
    \begin{equation}
        \op{H}(t) 
        = - \hbar \op{d}^\dagger \left[G_+(t) \op{b}^\dagger + G_-(t) \op{b} \right] + \text{H.c.}.\label{eq:H_RWA}
    \end{equation}
    Setting $G_+(t)=0$ reveals that the red-detuned drive $G_-(t)$ effectively cools the resonator by transferring excitations from the resonator to the cavity where they dissipate with rate $\kappa$. This is also known as optomechanical sideband cooling \cite{MarquardtPRL07, WilsonRaePRL07}. Similarly, it can be seen that the blue-detuned drive $G_+(t)$ effectively heats the resonator. Squeezing is generated through the interplay of both drives. To see this, one can introduce a Bogoliubov mode $\op{\beta}(r)$ with
    \begin{equation}
        \op{\beta}(r) = \op{b}\cosh{r} + \op{b}^\dagger\sinh{r},
        \label{eq:Bogoliubov}
    \end{equation}
    where the squeezing parameter $r(t)$ is defined by $\tanh{r(t)} = G_+(t)/G_-(t)$. The Hamiltonian can then be rewritten as
    \begin{equation}
        \op{H}(t) = -\hbar \mathcal{G}(t) \op{d}^\dagger \op{\beta}(r(t)) + \text{H.c.}
        \label{eq:H_Bogoliubov}
    \end{equation}
    with the effective coupling $\mathcal{G}(t) = \sqrt{G_-^2(t) - G_+^2(t)}$, where we have assumed $G_-(t), G_+(t) \in \mathbb{R}$ and without loss of generality $G_-(t) > G_+(t)$. For a fixed $r$, this Hamiltonian "cools" the resonator into the ground state of the Bogoliubov mode $\op{\beta}$. This ground state is a squeezed state where the variance of $\op{X}_1 = (\op{b}^\dagger + \op{b})/\sqrt{2}$ decreases exponentially with the squeezing parameter, $\Delta X_1^2 \propto \e{-2r}$. The resonator's state is considered to be squeezed if $\Delta X_1^2 < 1/2$. Due to the Heisenberg uncertainty principle, the variance of $\op{X}_2 = i(\op{b}^\dagger - \op{b})/\sqrt{2}$ increases as $\Delta X_2^2 \propto \e{2r}$. Based on these observations, Kronwald et al.~\cite{KronwaldPRA13} were able to show that one can drive the mechanical resonator into a squeezed steady state by using a protocol with constant amplitudes $G_\pm$. In the following, we use time-dependent amplitudes. Explicitly, this means that both the coupling $\mathcal{G}(t)$ and the squeezing parameter $r(t)$ are time-dependent due to the time-dependence of $G_+(t)$ and $G_-(t)$.


\subsection{Optimal control theory}
  \label{subsec:theory_OCT}

    For the design of the laser pulses, we use Krotov's method \cite{Krotov96, KonnovARC99, PalaoPRA03, ReichJCP12, GoerzNJP14, GoerzSP19}, a gradient-based optimization algorithm. In optimal control theory, the quantity to be optimized has to be incorporated into an appropriate optimization functional $J$. In each iteration, the gradient of this functional is then calculated to adjust the control pulses such that the functional is minimized. It usually consists of two parts,
    \begin{equation}
        J = J_T[\op{\rho}(T)] + \sum_l \int_0^T J_t[\op{\rho}(t), G_l(t)] \; \difd t,
        \label{eq:J_general}
    \end{equation}
    where $G_l(t)$ is the $l$-th time-dependent control field and $\op{\rho}(t)$ is the time-dependent state of the system. $J_T$ is the final-time functional and typically encodes the target of the optimization, \eg a specific state or in general a property one wants to achieve at final time $T$.
    
    The goal of the present work is to optimize squeezing. This corresponds to minimizing the variance of the resonator's quadrature $\op{X}_1$, so we choose the final time functional
    \begin{equation}
        J_T[\op{\rho}(T)] = \Delta X_1^2(T) = \Braket{\op{X}_1^2}(T) - \Braket{\op{X}_1}^2(T)\label{eq:def_functional}
    \end{equation}
    where $\braket{.}(T)$ denotes the expectation value with respect to the final state.

    Optimal control theory is often used to optimize a control protocol both in terms of its figure of merit $J_T[\op{\rho}(T)]$ and duration $T$. While with gradient-free optimization it is straightforward to minimize $T$ \cite{GoerzEPJQT15}, gradient-based optimizations have so far resorted to simply decreasing the final time until no more solution is found~\cite{CanevaPRL09}. Recently, a rescaling of time has been suggested to circumvent this issue in gradient-based methods~\cite{dasilva2024gate}. While a combination with Krotov's method should be possible, in the present work we have resorted to manual tuning of the protocol duration, similar to Refs.~\cite{CanevaPRL09,BasilewitschAQT19}. 

    In \cref{eq:J_general}, $J_t$ expresses time-dependent costs. It can be used, \eg to penalize high amplitudes of the control field or the population in unwanted states \cite{PalaoPRA08}. A common choice is \cite{PalaoPRA03}
    \begin{equation}
        J_t[\op{\rho}(t), G_l(t)] = \lambda_{a,l}\left[G_l(t) - G_{l,\text{ref}}(t) \right]^2.
    \end{equation}
    $G_{l,\text{ref}}(t)$ is typically chosen to be the field from the previous iteration, whereas $\lambda_{a,l}$ modulates the step size of the update.
    
    Starting from an initial guess, the pulses are iteratively optimized according to an update equation. In Krotov's method, the update equation is chosen such that monotonic convergence is ensured \cite{ReichJCP12}. Since Krotov's method assumes time-continuous controls, a suitable set of parameters $\lambda_{a,l}$ must be chosen such that the update applied in each iteration is not too large \cite{GoerzSP19, ReichJCP12}. For an open quantum system as considered here, the update equation for iteration step $k+1$ reads \cite{GoerzNJP14}
    \begin{align}
        G_l^{(k+1)}&(t)  
        = G_l^{(k)}(t) \nonumber\\
        &+ \frac{1}{\lambda_{a,l}} \mathfrak{Re}\left\{ \trace{\op{\chi}^{(k)}(t) \left. \del[\supop{L}]{G_l}\right|_{\substack{G_l^{(k+1)}(t)}} \op{\rho}^{(k+1)}(t) }\right\},
        \label{eq:update_epsilon}
    \end{align}
    where $\del[\supop{L}]{G_l}$ is the derivative of the Liouvillian with respect to the $l$-th control.
    $\op{\chi}^{(k)}(t)$, usually referred to as the costate, is evolved backwards in time according to the adjoint Liouvillian~$\supop{L}^\dagger$ with the control fields of the $k$-th iteration,
    \begin{equation}
        \deriv{t} \op{\chi}^{(k)}(t) = -\supop{L}^\dagger\left[\left\{G_l^{(k)}(t)\right\}\right]\op{\chi}^{(k)}(t)
        \label{eq:motion_costates}
    \end{equation}
    with boundary condition
    \begin{equation}
        \op{\chi}^{(k)}(T) = - \nabla_{\op{\rho}} J_T\Bigr|_{\substack{\op{\rho}^{(k)}(T)}}.
    \end{equation}
    The states $\op{\rho}^{(k+1)}(t)$ are obtained by solving the master equation \eqref{eq:Lindblad-Master} with the new set of pulses $\{G_l^{(k+1)}(t)\}$
    \begin{equation}
        \deriv{t} \op{\rho}^{(k+1)}(t) = \supop{L} \left[\left\{G_l^{(k+1)}(t)\right\}\right]\op{\rho}^{(k+1)}(t)
        \label{eq:motion_rho}
    \end{equation}
    with the initial condition
    \begin{equation}
        \op{\rho}^{(k+1)}(0) = \op{\rho}_0,
    \end{equation}
    in which $\op{\rho}_0$ is the initial state of the system.


\section{Squeezing generation using optimal control}
  \label{sec:results}

    In the following, we show that one can speed up the transition into a squeezed state by using time-dependent amplitudes. Reference~\cite{BasilewitschAQT19} already reported that a protocol with time-dependent amplitudes can speed up the squeezing process when optimizing towards a specific squeezed target state. We show that achieving speedups is also possible when optimizing for arbitrary states of high squeezing, rather than for a specific state. In contrast to Ref.~\cite{BasilewitschAQT19}, we investigate the quantum speed limit for squeezing and how it is influenced by the different system parameters. We furthermore study the influence of the counterrotating terms in \cref{eq:H_final_with_rotating_terms} on the control solutions.


  \subsection{Details of the optimization}
    
    The protocol starts with the cavity initially in the vacuum state and the resonator in a thermal state with thermal occupancy $n_\mathrm{th}$. The goal of the optimization is to find pulses that transform the resonator's initial thermal state into a highly squeezed one. To accomplish this, we employ Krotov's method as presented in \cref{subsec:theory_OCT}.

    Throughout this work, we use parameters from the experiment by Wollman et al.~\cite{WollmanS15} (see \cref{tab:parameters}), which realized the setup proposed in Ref.~\cite{KronwaldPRA13}. However, we use a temperature of about 0.4~mK for the thermal bath, compared to 10~mK in the experiment, in order to ease the numerical simulation of the system dynamics. Larger values of $n_\mathrm{th}$, \ie higher temperatures, result in higher initial population of the resonator and thus also a larger system size that needs to be simulated. We will argue below that the temperature does not influence the control strategy and we will also discuss how the quantum speed limit is affected by temperature.
    \begin{table}[t]
        \centering
        \caption{System parameters used in all optimizations, taken from the experiment in Ref.~\cite{WollmanS15}.}
        \begin{tabular}{l|c|l}\hline\hline
            Cavity frequency     & $\omega_\text{cav}$ & $2\pi\times6.23$ GHz \\
            Resonator frequency  & $\Omega$            & $2\pi\times3.6$ MHz  \\
            Coupling strength    & $g_0$               & $2\pi\times36$ Hz    \\
            Cavity decay rate    & $\kappa$            & $2\pi\times450$ kHz  \\
            Resonator decay rate & $\Gamma$            & $2\pi\times3$ Hz     \\
            Thermal occupancy    & $n_\mathrm{th}$     & $2$                  \\\hline\hline
        \end{tabular}
        \label{tab:parameters}
    \end{table}

    In experiments, microwave fields can be manipulated on timescales down to subnanoseconds \cite{YaoOC11a,HaberlePRL13}. Since our simulations employ time steps larger than that, we do not have to account for finite ramping times such that the pulses can start at arbitrary values in the simulations.
    As guess pulses we use constant amplitudes with $G_-/2\pi = 5.8$~kHz and $G_+/G_- = 0.7$, similar to Ref.~\cite{BasilewitschAQT19}, and take the amplitudes to be real. This implies squeezing a certain quadrature of the mechanical motion (in the interaction picture), \ie a specific orientation of the squeezed state in phase space. It can readily be generalized towards arbitrary orientations by adding a constant complex phase to the pulse amplitudes.

    To simplify and speed up the optimizations, we model the system dynamics in the RWA, \ie with the Hamiltonian in \cref{eq:H_RWA}.
    In general, the effects of the counterrotating terms in \cref{eq:H_final_with_rotating_terms} are only negligible if one can separate the timescales of the mechanical motion and the cavity decay, \ie if $\kappa \ll \Omega$. In the literature, this is also referred to as the good cavity limit or the resolved sideband regime \cite{AspelmeyerRMP14, Clerk20}. 
    As we use system parameters for which $\kappa/\Omega = 1/8$, it is possible that the counterrotating terms impact the time evolution, particularly at high amplitudes \cite{KronwaldPRA13}. Therefore, special attention has to be paid to whether the control strategy relies on the RWA to achieve higher squeezing. Thus, we only use the RWA in the optimizations to save resources but not in the final propagation, and compare the time evolution with and without RWA to quantify the effect of it. Based on this comparison, we manually adjust the pulses to ensure they are physically meaningful. The results we show in the following always refer to the adjusted pulses.


  \subsection{Optimization results}\label{subsec:results_dynamics}

    \begin{figure}[tb]
        \centering
        \includegraphics[width=\linewidth]{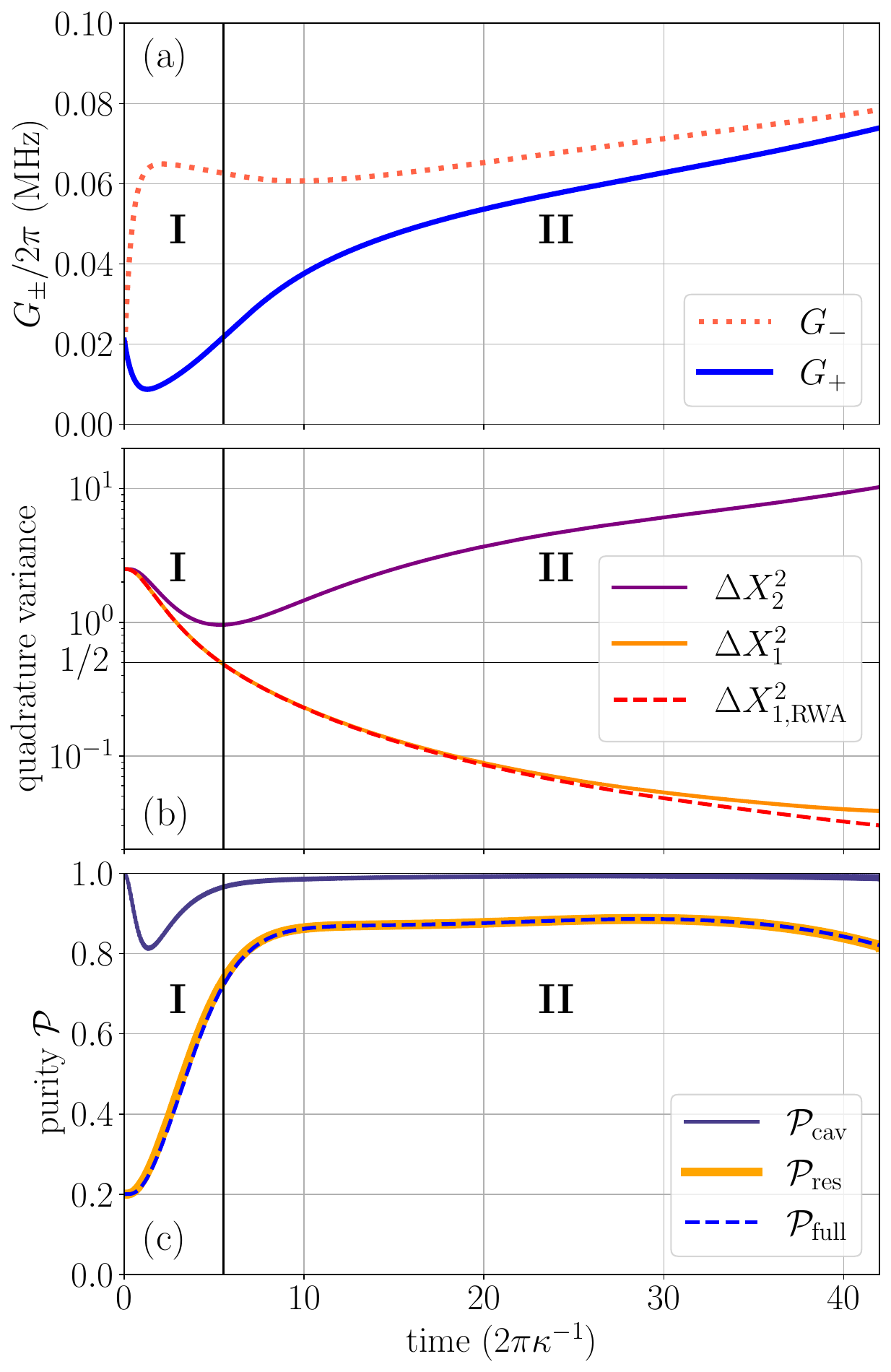}
        \caption{Example for the outcome of an optimization taking $T=42 \times 2\pi\kappa^{-1} \approx 93.3~\mu$s. (a) Optimized pulse amplitudes. (b) Variances of both quadratures of the resonator without RWA (solid lines), and with RWA (dashed line). The line for $\op{X}_2$ coincides with the one without RWA. (c) Purity $\mathcal{P}=\trace{\op{\rho}^2}$ of the resonator, the cavity and the full system. Maximum squeezing (variance divided by the zero-point fluctuation of $1/2$) with RWA: 12.2~dB, without: 11.1~dB}
        \label{fig:results_42kappa}
    \end{figure}
    We have conducted optimizations for final times between $T=0.1~\mu$s and $T=150~\mu$s and identified two different regimes for the optimized pulses. For optimizations, in which the final time $T>2\pi \kappa^{-1} \approx 2.2~\mu$s, large amounts of squeezing are achieved,  and a general control strategy can be identified. For optimizations with $T<2\pi \kappa^{-1}$, only small amounts of squeezing can be achieved and there is no control strategy easily identified. In the next section, we argue that this is closely connected to the quantum speed limit of squeezing. Since the explanation is based on the control strategy for the larger final times, we discuss it first.
    
    The control strategy consists of two parts -- the cooling phase (I) and  the squeezing phase (II). An exemplary optimization is shown in \figref[]{fig:results_42kappa}. The cooling phase (I) lasts to about $t=T/8$ and reduces both variances of the resonator close to the zero-point fluctuation of $1/2$ (see \figref[(b)]{fig:results_42kappa}), \ie the resonator is cooled from its thermal state to a state close to the ground state. To accomplish this, the red drive’s amplitude $G_-$ which cools the resonator is increased while $G_+$, the heating blue drive’s amplitude, is kept low, see \figref[(a)]{fig:results_42kappa}. \figureref[(c)]{fig:results_42kappa} shows the purity of the full composite system state and the purity of the reduced states of both the resonator and the cavity as a function of time. During the cooling phase, the purity of the resonator's state significantly increases while the one of the cavity briefly decreases. This indicates that during this phase, the initial thermal population in the resonator transfers to the cavity, where it dissipates at a rate $\kappa$.
    During the squeezing phase (II), both amplitudes are increased while their difference decreases over time. This can be explained using the Bogoliubov transformation, \cf\cref{eq:Bogoliubov,eq:H_Bogoliubov}. By decreasing the difference between $G_+$ and $G_-$, the ratio of the two approaches one, meaning the squeezing parameter $r$ increases. Consequently, the $\op{X}_1$-quadrature is squeezed, while the variance of $\op{X}_2$ increases. On the other hand, the effective coupling $\mathcal{G}$ decreases as the amplitudes become closer in value. However, $\mathcal{G}$ is dependent on the absolute difference and can thus be maintained at a higher level by also increasing the red drive's amplitude.

    To check the performance of the pulses optimized in the RWA, we compare the time evolution of the variance of $\op{X}_1$ with and without RWA in \figref[(b)]{fig:results_42kappa}. One can see that the discrepancy between the variance with and without RWA is very small. It is thus reasonable to conduct the optimizations in the RWA, since this allows us to find physically meaningful solutions realizing large squeezing  while at the same time reducing the computational effort.

    To quantify the speedup gained from the optimization, we compare the obtained solution to a protocol with fixed amplitudes, as it was originally proposed in Ref.~\cite{KronwaldPRA13}. Therefore, \figref[(a)]{fig:protocol_comp_42kappa} displays the optimized pulses from \figref[(a)]{fig:results_42kappa} (dash-dotted line) together with the constant-amplitude protocol (solid line). As can be seen in \figref[(a)]{fig:results_42kappa}, the average of $G_-/2\pi$ for the optimized pulse is about 0.07~MHz. Thus, to be able to compare the results in terms of the used resources, we also set for the constant protocol $G_-/2\pi=0.07$~MHz. The value of $G_+$ that achieves the highest squeezing in the given time can be found by a simple line search and turns out to be $G_+=0.86\,G_-$. The resulting time evolution of the variances $\Delta X_1^2$ and $\Delta X_2^2$ is shown in \figref[(b)]{fig:protocol_comp_42kappa}. First of all, the optimized solution achieves in total a larger squeezing of about 11.1~dB compared to the 10.1~dB of the constant protocol. Secondly, it reaches the value of 10.1~dB within a time of about $32\times2\pi\kappa^{-1}$, which is only $76\%$ of $42\times2\pi\kappa^{-1}$, the time the constant protocol needs.
    On the other hand, one can also ask how fast a squeezing of 11.1~dB can be achieved with the constant protocol and the amplitude fixed to $G_-/2\pi=0.07$~MHz. The answer is $51\times2\pi\kappa^{-1}$ with a ratio of $G_+/G_-\approx0.90$. Also in this comparison the optimized protocol is faster and needs only about $82\%$ of the constant protocol's duration.

    The speedup can be understood in terms of the control strategy identified above. In the constant protocol, the cooling phase is skipped and squeezing starts immediately. Since the amplitudes $G_-$ and $G_+$ must be close for high squeezing, the coupling of the two systems is low and the entire squeezing slow, especially when driving the system towards a highly squeezed state. The main reason for the speedup is thus the cooling phase in the beginning, in which the effective coupling $\mathcal{G}$ is large allowing for the initial thermal population in the resonator to be removed efficiently. A further small speedup is gained by increasing $G_-$ towards the end of the protocol, \ie investing more of the available resources towards the end where $\mathcal{G}$ becomes smaller.
    In general, we observe for all protocol durations that one can reduce the time to achieve a certain amount of squeezing by $15\%$ to $25\%$ with the optimized pulses compared to the constant protocol.

    It is interesting to compare our optimized protocols to a recent proposal for squeezing a mechanical mode via detuning-switched driving \cite{LiPRA2023}. The proposal uses only a single drive with effective amplitude $G$. It consists in the periodic application of short pulses with strongly increasing detuning. As a result an effective linear force  acts on the resonator, squeezing it. This comes with the advantage of not being limited by the cavity decay rate such that in principle squeezing can potentially be generated even faster than in our protocols. However, Ref.~\cite{LiPRA2023} operates in the strong-coupling regime $G \gg \Omega$ and requires a precise modulation of the laser power and frequency on timescales $t_0\ll2\pi G^{-1}$. In contrast, our optimized protocol just requires a slow modulation of the amplitudes and works for smaller amplitudes. It is thus potentially easier to implement in an experiment.

    \begin{figure}[t]
        \centering
        \includegraphics[width=\linewidth]{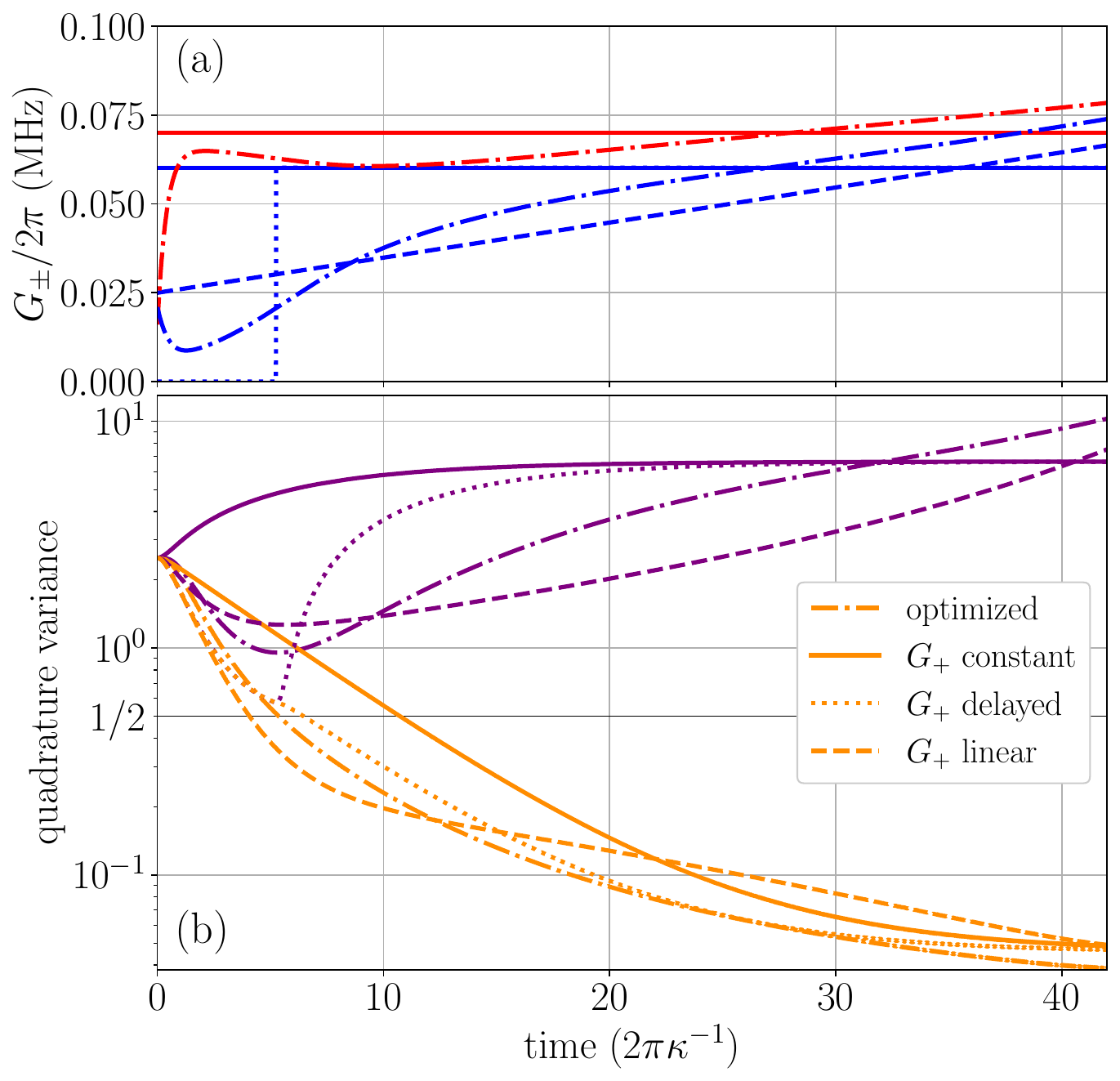}
        \caption{Comparison of different squeezing protocols. (a) Pulse amplitudes ($G_-$ in red and $G_+$ in blue) for the optimized protocol from \figref[(a)]{fig:results_42kappa} (dash-dotted) in comparison with the constant protocol from Ref.~\cite{KronwaldPRA13} (solid) and two proposals for a simplified squeezing protocol based on the control strategy found from the optimized pulses -- a linearly increasing protocol (dashed), and a constant-amplitude protocol with a delayed blue pulse $G_+$ (dotted). $G_+(T)/G_-(T)$ is 0.86 for the constant and delayed protocol; and $0.95$ for the linear protocol. (b) The resulting temporal evolution of the variances of the resonator's quadratures without RWA ($\Delta X_1^2$ in orange and $\Delta X_2^2$ in purple). Maximum squeezing for the optimized pulses: 11.1~dB; for the constant protocol from Ref.~\cite{KronwaldPRA13}: 10.1~dB; for the delayed protocol: 10.3~dB; for the linear protocol: 10.1~dB.}
        \label{fig:protocol_comp_42kappa}
    \end{figure}

    The two-stage control strategy and the form of the pulses in \figref[(a)]{fig:results_42kappa} suggest that deriving an even simpler protocol is possible. Indeed, the red drive's amplitude is almost constant during the protocol, whereas that of the blue drive is low during the cooling phase and increased during the squeezing phase. Therefore, \figref[(a)]{fig:protocol_comp_42kappa} also displays two possible simplified squeezing protocols. The first one (dashed line) consists of a constant red drive combined with a linear ramp in the blue drive, whereas the second (dotted line) is a protocol with constant amplitudes where the blue drive is switched on with a time delay. To be able to compare the results, we set for both new protocols $G_-/2\pi=0.07$~MHz, similarly to before. In the linear-ramp protocol, we let $G_+$ increase linearly from the initial value $G_+/2\pi\approx0.025$~MHz to a final ratio $G_+(T)/G_-(T)$ which we take $\approx0.95$ as the ratio found in the optimized pulses, see \figref[(a)]{fig:results_42kappa}. In the time-delay protocol, the blue pulse is switched on  at the time at which the variance of $\op{X}_1$ becomes equal to the zero-point fluctuation in the optimized squeezing scheme. The optimal amplitude of the blue drive is again determined by a line search. \figureref[(b)]{fig:protocol_comp_42kappa} shows the quadrature variances over time for the constant protocol, the optimized protocol, and the two simplified protocols. One can see that the $\op{X}_1$-quadratures of both simplified protocols reach the zero-point fluctuations at almost the same time as the one in the optimized protocol does. This is because they are both able to leverage the stronger interaction strength of the two subsystems during the cooling phase.
    After the cooling phase, the time-delay protocol (dotted lines) performs almost as well as the optimized one (dash-dotted lines), but eventually the variance reaches a plateau as the resonator approaches its steady state, similar to the constant protocol (solid lines). 
    The linear-ramp protocol (dashed lines) initially outperforms the protocol with constant amplitudes. However, it becomes worse at later times and only catches up in the end when the constant one runs into the steady state.
    \begin{figure}[t]
        \centering
        \includegraphics[width=\linewidth]{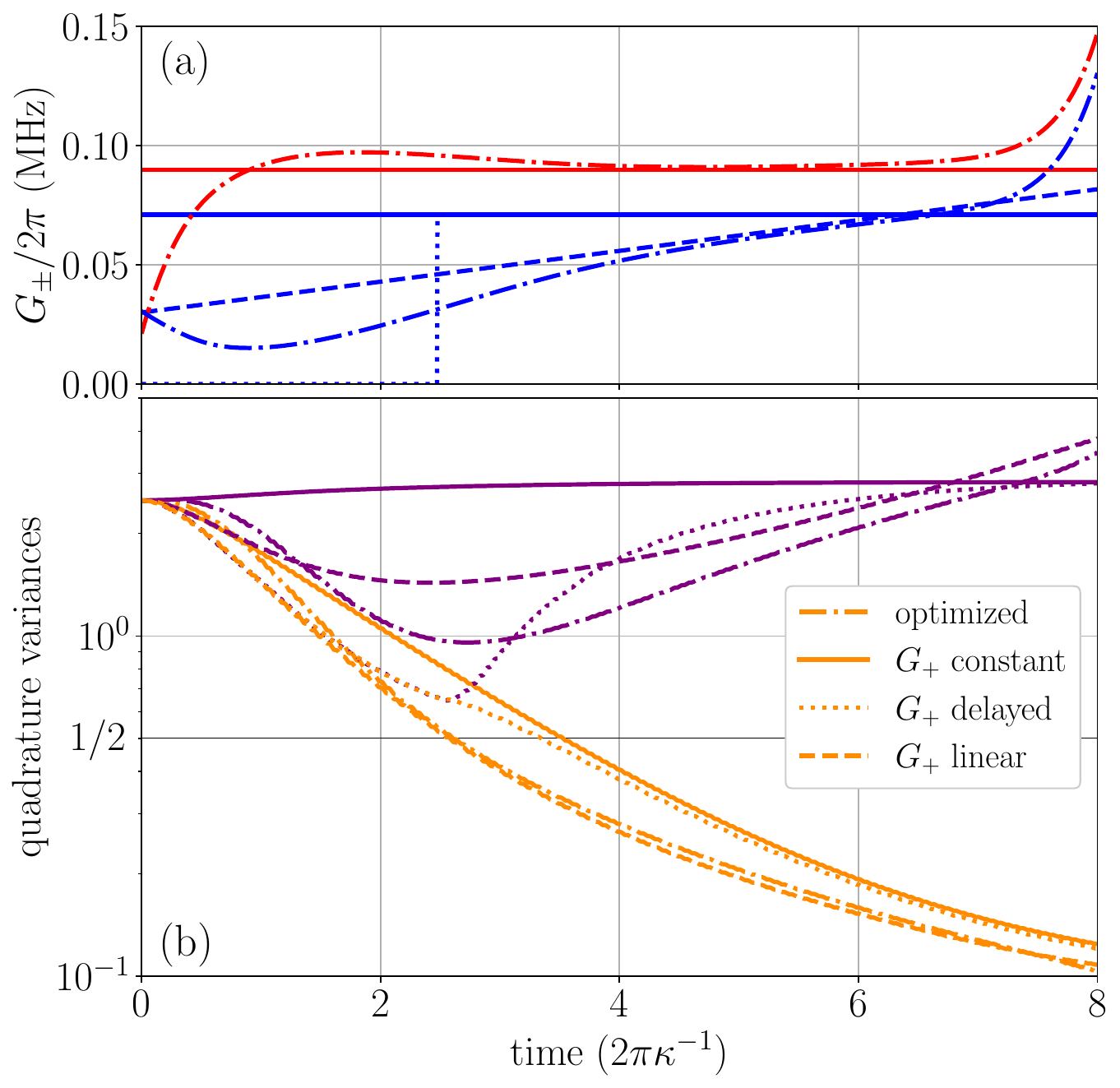}
        \caption{Comparison of different squeezing protocols, analogously to \figref[]{fig:protocol_comp_42kappa} but for a shorter protocol duration. (a) Pulse amplitudes. $G_+(T)/G_-(T)$ is $0.70$ for the constant protocol from Ref.~\cite{KronwaldPRA13}; $0.71$ for the delayed protocol; and $0.91$ for the linear protocol. (b) The resulting temporal evolution of the variances of the resonator's quadratures (without RWA). The achieved maximum squeezing is 6.8~dB for the optimized pulses; 6.0~dB for the constant protocol from Ref.~\cite{KronwaldPRA13}; 6.2~dB for the delayed protocol; and 6.6~dB for the linear protocol.}
        \label{fig:protocol_comp_8kappa}
    \end{figure}
    The linear-ramp protocol is most advantageous for shorter protocol durations.
    This is illustrated in \figref[]{fig:protocol_comp_8kappa} showing again the four different protocols, but this time for a duration of $T=8\times2\pi\kappa^{-1}\approx18~\mu$s (with the parameters of the different protocols chosen in the same way as in \figref[]{fig:protocol_comp_42kappa}). Again, the optimized protocol performs better than the constant protocol, but for this short duration the linear-ramp protocol performs almost as good as the optimized one and the time-delay protocol is only slightly faster than the constant one. Thus, for the two simplified protocols, the time-delay protocol is  better for longer protocol durations and the linear protocol for shorter durations. This can be explained as follows:
    For shorter durations, it becomes more important to remove the initial thermal energy in the resonator fast, also for the constant protocol. Thus, it is important to have a large effective coupling $\mathcal{G}=\sqrt{G_-^2 - G_+^2}$ throughout the protocol. This is achieved by lowering the blue drive's amplitude, at the expense of squeezing, and explains the optimal ratio $G_+/G_-=0.70$ for the constant protocol in \figref{fig:protocol_comp_8kappa}, compared to $G_+/G_-=0.86$ in \figref{fig:protocol_comp_42kappa}. It also means that the speedup that is gained in the extra cooling phase of the time-delay protocol is less significant. On the other hand, the cooling phase takes a larger part of the overall protocol. For the linear-ramp protocol, the blue drive has thus already been increased to moderate values at the end of the cooling phase. In other words, for short durations, the linear-ramp protocol realizes an optimal trade-off between cooling (via the effective coupling $\mathcal{G}$) and squeezing.
    For longer protocols, the opposite is the case. 
    Here, the relative duration of the cooling phase is shorter and it is more important to realize large values of the squeezing parameter $r=\text{artanh}(G_+/G_-)$ with a lower effective coupling $\mathcal{G}$. This means that in the time-delay protocol, the relative increase of $\mathcal{G}$ during the cooling phase compared to the constant protocol is larger, which leads to greater speedup.
    On the other hand, in the linear-ramp protocol, the blue drive's amplitude grows more slowly and while the protocol benefits initially from a large effective coupling $\mathcal{G}$, it cannot reach the largest amount of squeezing since the squeezing parameter $r$ is small for a large part of the protocol.

    To summarize, both simplified protocols can achieve speedups comparable to the optimized one, with the linear-ramp protocol performing better for shorter protocol durations and the time-delay protocol for longer ones. Due to their very simple pulse shapes, they are potentially easier to implement in an experiment. Furthermore, they provide a straightforward way to generalize the two-stage control strategy to protocols of arbitrary duration without the need of reoptimization.


  \subsection{Quantum speed limit for squeezing}\label{subsec:QSL}

    With the control strategy for the larger final times at hand, we can rationalize why it breaks down for shorter times, and thereby determine the quantum speed limit for squeezing. In order to quantify the quantum speed limit, we need to define a threshold for squeezing.

    In optomechanical systems, squeezing up to one half of the zero-point fluctuation can be readily achieved by a simple parametric driving scheme \cite{MilburnOC81, MariPRL09, WoolleyPRA08, NunnenkampPRA10, LiaoPRA11} while squeezing beyond this value requires more advanced techniques such as QND measurements combined with a coherent feedback operation \cite{BraginskyS80, RuskovPRB05, ClerkNJP08, Clerk20}. This suggests to take one half of the zero-point-fluctuation as a threshold for the quantum speed limit. Following the literature \cite{KronwaldPRA13, VinantePRL13, QinPRL2022}, we will refer to it in the following as the 3~dB limit.

\begin{figure}[t]
        \centering
        \includegraphics[width=\linewidth]{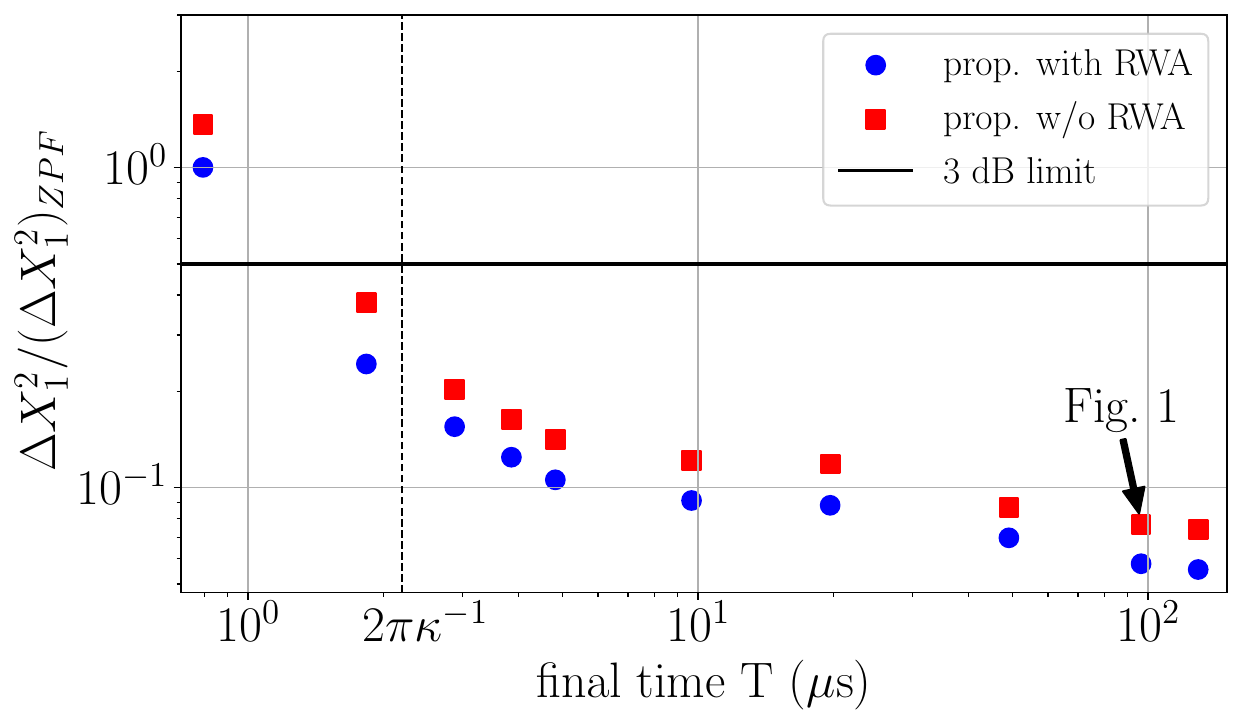}
        \caption{Maximum amount of squeezing for different final times. Shown as blue dots are the results obtained in the RWA, red squares show the maximum amount of squeezing obtained with the same pulses propagated without the RWA. The obtainable squeezing is significantly reduced for final times $T \lesssim 2\pi\kappa^{-1}$.
        Interestingly, it seems that removing the RWA increases the variance always by a constant factor.}
        \label{fig:QSL}
    \end{figure}

    \figureref{fig:QSL} shows the maximum amount of squeezing achieved in each optimization for a given final time. Squeezing increases (\ie the variance decreases) monotonously with longer protocol durations as expected. Further, one can see that for final times $T$ in the range of the decay time $2\pi\kappa^{-1}\approx2.2~\mu$s or below, squeezing rapidly decreases. It should be noted that the average amplitude of both drives also increases towards shorter protocol durations.
    The reason for this becomes apparent when considering the control strategy identified in \cref{subsec:results_dynamics}. Since the cooling phase depends on the decay, the population always takes a time proportional to $2\pi\kappa^{-1}$ to dissipate. This implies that in general a smaller protocol duration results in less time for the squeezing phase.
    Consequently, to achieve the same amount of squeezing, it is necessary to increase both amplitudes during the squeezing phase more.
    However, care has to be taken, since then the counterrotating terms become more relevant and the RWA is not a good approximation anymore.
    This is due to the variation of the pulse amplitude derivatives over multiple periods of the oscillation which prevents the counterrotating terms to average to zero. This is important because Ref.~\cite{KronwaldPRA13} has shown that the counterrotating terms effectively heat the Bogoliubov mode $\op{\beta}$, which corresponds to a decrease in squeezing. This decrease is also visible in \figref{fig:QSL}. For each final time we show the maximum squeezing that is reached when propagating the system in the RWA and without it. 
    One can see that the ratio of the variances without and with RWA is always around $\sim1.4$.
    This means that for smaller final times where the pulse amplitudes are larger and change more rapidly, the absolute difference between the two cases is greater, which supports the discussion above.
    This poses a physical limit to how large one can make the pulse amplitudes before the squeezing protocol breaks down. By that, also the amount of squeezing that can be achieved in a given time is limited, which explains why the squeezing decreases from $T=150~\mu$s to $T=10~\mu$s.

    The rapid decrease in squeezing for protocol durations below $T=10~\mu$s is explained in a similar sense. For times $T\lesssim2\pi\kappa^{-1}$, the cooling phase cannot be sufficiently long and the thermal population of the resonator transfers to the cavity but remains there. Then, the control strategy breaks down and squeezing beyond the $3$~dB limit cannot be achieved anymore. This poses a fundamental bound on how fast squeezing can be achieved --- the quantum speed limit. 
    Note that our finding is different from the results of Ref.~\cite{BasilewitschAQT19} targeting specific (squeezed) states, where 
    the maximum and average amplitude of the pulses turned out to be monotonically decreasing functions of the protocol duration and follow a power-law dependence. While we find a similar dependence for the pulse amplitudes, we have not considered specific target states and can thus identify the physical limit for squeezing in the system. This is further corroborated by the protocol durations which in Ref.~\cite{BasilewitschAQT19} were more than an order of magnitude larger than the cavity decay time, and the moderate amplitude values. This explains why neither the breakdown of the control strategy nor the limiting effect of the counterrotating terms were observed in Ref.~\cite{BasilewitschAQT19}.

    This explanation allows us to analyze the role of the system parameters -- the thermal occupancy $n_\mathrm{th}$, the resonator frequency $\Omega$, and the decay rates of the cavity, $\kappa$, and the resonator, $\Gamma$ -- in the squeezing process and the control protocol. We start by discussing the effect of $n_\mathrm{th}$.
    We do not expect the overall control strategy to change for a higher thermal occupancy, since the way to deal with the increased initial population is to increase the amplitude of the red drive during the cooling phase (as far as possible within the rotating wave approximation).
    Alternatively, one could extend the cooling phase. For instance, one could start with an additional cooling phase, wherein $G_+$ is set to 0, in order to cool the resonator down to $n_\mathrm{th}=2$ prior to commencing with our control protocol. Thus, we generally expect the quantum speed limit to shift to longer times for higher temperatures.

    The resonator frequency $\Omega$ only affects squeezing through the influence of the counterrotating terms. As already demonstrated in Ref.~\cite{KronwaldPRA13}, these terms limit squeezing and their impact grows as one moves away from the good cavity limit $\kappa \ll \Omega$. Thus, we generally expect that smaller values of $\kappa/\Omega$ will result in higher possible squeezing and a shorter time for the quantum speed limit.

    Further, the control scheme and the quantum speed limit are inherently connected to the decay time $2\pi\kappa^{-1}$. For lower rates $\kappa$, the duration of the cooling phase must be increased because the thermal population is now removed from the system more slowly. In contrast to a higher temperature, this cannot be overcome by means of higher amplitudes.

    Since we only consider timescales $T \ll 2\pi\Gamma^{-1}\approx0.33$~s, the resonator decay rate is basically irrelevant in our optimizations. However, if one considers higher values for $\Gamma$ or steady-state squeezing \cite{KronwaldPRA13, BasilewitschAQT19}, it is also a factor that limits squeezing, since it leads to a decay of the resonator back towards the initial thermal state. This decay must be counteracted to achieve the same results, which requires additional resources. However, as long as $\Gamma$ stays orders of magnitudes smaller than $\kappa$ and $\Omega$, we do not expect it to influence the quantum speed limit.

    \begin{figure}[t]
        \centering
        \includegraphics[width=\linewidth]{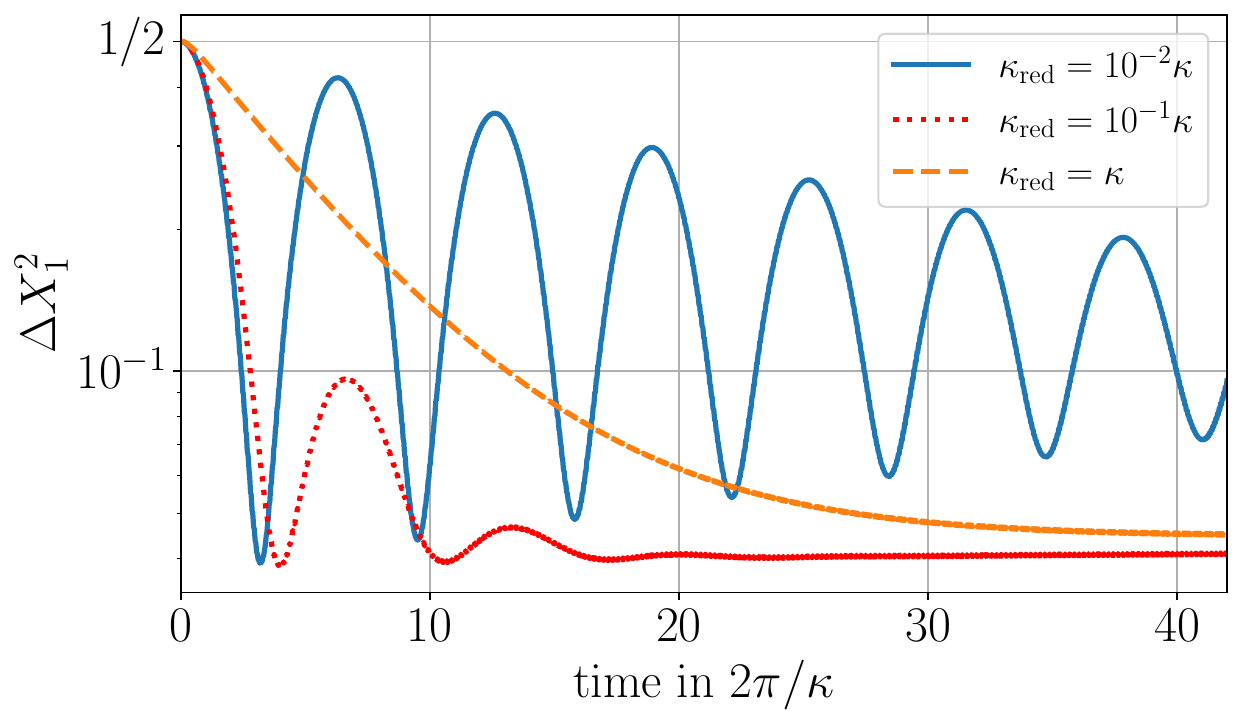}
        \caption{Variance of $\op{X}_1$ for the constant amplitudes shown in \figref[(a)]{fig:protocol_comp_42kappa} for different artificially reduced cavity decay rates $\kappa_\text{red}$ and a thermal occupancy of the resonator of $n_\text{th}=0$. The time is given in units of the decay rate $\kappa=450$~kHz used in the rest of this paper.}
        \label{fig:variance_kappa-comp_const}
    \end{figure}

    The cavity decay greatly enhances the speed of the squeezing process, since it allows to remove the entropy initially stored inside the resonator from the system during the cooling phase. It is natural to ask whether the decay is also helpful during the squeezing phase, \ie whether the protocol also benefits from a high cavity decay rate $\kappa$ with $n_\mathrm{th}=0$. Therefore, we simulate the system dynamics with the constant controls shown in \figref[(a)]{fig:protocol_comp_42kappa}, artificially reducing the decay rates by one, resp. two orders of magnitude and a thermal population of the resonator $n_\mathrm{th}=0$. The resulting time evolutions of the variance $\Delta X_1^2$ are shown in \figref{fig:variance_kappa-comp_const}. They all approach a different steady-state value where the ones for lower decay rates also show damped oscillations and reach the maximum squeezing much earlier than for $\kappa_\text{red}=\kappa$. Neglecting again the effect of $\Gamma$, the results can be explained as follows.
    While for $\kappa_\text{red}=\kappa$ the cavity acts like a reservoir to the resonator, the time evolution becomes almost unitary in the limit of $\kappa_\text{red} \rightarrow 0$. Without dissipation, and since the Hamiltonian  \eqref{eq:H_final_with_rotating_terms} is symmetric,
    the reduced states of resonator and cavity are identical at all times, which means that the cavity evolves into a squeezed state as well. At later times, recurrences occur, which partially destroy the squeezing that was built up at previous times. This is the reason for the oscillatory behavior seen for the reduced decay rates.
    For the squeezing phase, a lower decay rate is thus more beneficial, since the transition to the squeezed state is only governed by the amplitude of the drives and thus much faster. Note that a smaller decay rate also allows in principle for a more highly squeezed state when choosing a larger ratio $G_+/G_-$. However, smaller dissipation also implies a longer cooling phase, since the thermal energy is removed from the system more slowly. One thus needs to balance cooling and squeezing when choosing a suitable decay rate for the system at hand.
    
    To summarize, a high cavity decay rate is advantageous for achieving squeezed states quickly, since it allows to remove the thermal energy inside the resonator efficiently. On the other hand, if one aims at a highly squeezed state and time is less important, a lower cavity decay rate might be more favorable.


\section{Conclusion and Outlook}
    \label{sec:conclusion}

    We have investigated how to generate squeezing in an optomechanical system in a limited amount of time. To this end, we have considered driving at both mechanical sidebands, originally proposed in Ref.~\cite{KronwaldPRA13}. Building on previous work \cite{BasilewitschAQT19}, we have taken the amplitudes to be time-dependent and used optimal control theory to design suitable time-dependencies. Contrary to Ref.~\cite{BasilewitschAQT19}, here we have used the variance of one of the quadratures of the mechanical motion as the target functional to directly optimize squeezing. This has allowed us to identify a different class of control solutions.

    The optimized pulse shapes are easily rationalized: The control strategy for protocols with a duration longer than the decay time of the cavity consists of two phases. It starts with a cooling phase in which the initial thermal population of the resonator dissipates via the cavity, followed by a phase in which the state of the resonator is driven into a squeezed state.
    The protocol achieves a speedup of about $15\%$ to $25\%$ compared to a control scheme with constant amplitudes \cite{KronwaldPRA13}. The simulations were done using parameters from an actual experiment \cite{WollmanS15} (except for temperature) and are thus close to physical reality. The obtained pulse shapes could for example be implemented with arbitrary waveform generators. By extending the duration of the cooling phase, the protocol can also be applied to systems with higher initial thermal populations. A recent proposal for squeezing a mechanical mode via detuning-switched driving \cite{LiPRA2023} could potentially achieve even faster speedups, but it operates in the strong-coupling regime and requires fast modulation of the laser power and frequency. In contrast, our scheme is readily applicable for moderate amplitude values and slow amplitude modulations.

    Based on the insight into the control strategy, we have also derived two simplified protocols. In both protocols, the amplitude of the red-detuned drive is held constant. In the first simplified protocol, the blue-detuned drive is ramped up linearly, whereas in the second one it is initially zero and switched on with a time delay. These protocols achieve speedups comparable to those obtained in the optimized protocols, and may be even easier to implement in an experiment. Moreover, they provide a simple way to implement the two-stage control strategy for arbitrary protocol duration, without the need of reoptimization.

    Finally, we have determined the quantum speed limit for squeezing by carrying out optimizations for various protocol durations. Our control strategy utilizes the dissipation of the cavity and thus benefits from higher cavity decay rates (as long as one stays within the good cavity limit). For protocol durations shorter than the cavity decay time, the control strategy breaks down and the amount of squeezing that can be achieved is strongly reduced. This implies that the quantum speed limit is inherently connected to the decay time of the cavity. We have also found, similarly to Ref.~\cite{KronwaldPRA13}, that the counter-rotating terms limit squeezing. As a result, larger amounts of squeezing cannot simply be achieved by increasing the amplitude of the drives.

    Our work provides an example for maximizing an observable of interest directly with quantum optimal control, similar to earlier work minimizing energy~\cite{DoriaPRL2011}. Targeting the actual quantity of interest -- the amount of squeezing -- rather than a specific state (that happens to be squeezed)~\cite{BasilewitschAQT19} allows for more flexibility in the optimization which in turn has resulted in identifying different control solutions. Our work thus underlines the importance of choosing the most suitable figure of merit in quantum optimal control.


\section*{Acknowledgements}

Funding from the Deutsche Forschungsgemeinschaft (DFG, German Research Foundation) – Projektnummer 277101999 – TRR 183 (project B02)
and from the German Federal Ministry of Education and Research (BMBF) within the project NiQ (13N16201) is gratefully acknowledged.


\appendix
\section{Derivation of the optomechanical Hamiltonian}\label{appendix:derivation}

    For completeness, we show how to derive the Hamiltonian in \cref{eq:H_final_with_rotating_terms} reported in Ref.~\cite{KronwaldPRA13}, paying special attention to new terms that arise due to the finite derivatives of the pulse amplitudes.
    We start by considering the Hamiltonian in \cref{eq:H_with_H_drive} and apply a displacement transformation with $\op{d} \rightarrow \bar{a}_+(t) \e{-i\omega_+t} + \bar{a}_-(t) \e{-i\omega_- t} + \op{d}$. This transformation can be executed by means of a displacement operator $\op{D}(\alpha)~=~ \e{\alpha \op{d}^\dagger - \alpha^*\op{d}}$, which has the property $\op{D}^\dagger(\alpha)\op{d}\op{D}(\alpha) = \alpha + \op{d}$ \cite{Walls08}. The Hamiltonian is then transformed using the standard transformation rules
    \begin{align}
        \op{H}_\text{disp}(t) = \op{D}^\dagger(\alpha(t)) & \op{H}(t) \op{D}(\alpha(t))\nonumber\\
        &+ i\hbar \left(\deriv{t}\op{D}^\dagger(\alpha(t))\right) \op{D}(\alpha(t)),
        \label{eq:app_H_displ}
    \end{align}
    where $\alpha(t) = \bar{a}_+(t) \e{-i\omega_+t} + \bar{a}_-(t) \e{-i\omega_- t}$.
    This transformation removes the driving Hamiltonian, \cref{eq:H_dr}. After neglecting terms proportional to the identity and going into an interaction picture with respect to the free Hamiltonian (the first two terms in \cref{eq:H_with_H_drive}), the Hamiltonian looks like
    \begin{alignat}{2}
        \op{H}_\text{disp}' = 
        &- \hbar \op{d}^\dagger \left[G_+ \op{b}^\dagger + G_- \op{b} 
        \right] & &+ \text{H.c.} \nonumber\\
        &- \hbar \op{d}^\dagger \left[G_+ \op{b}
        e^{-i2\Omega t} + G_- \op{b}^\dagger e^{i2\Omega t}\right] & &+ \text{H.c.}\nonumber\\
        &- \frac{i\hbar}{g_0} \op{d}^\dagger\left[\deriv[G_+]{t} e^{-i\Omega t} + \deriv[G_-]{t} e^{i\Omega t}\right] 
        & &+ \text{H.c.} \nonumber\\
        &-\frac{\hbar}{g_0} \left[G_+^2 + G_-^2 + 2 G_+G_- \cos{2\Omega t}\right] \op{b}^\dagger e^{i\Omega t} & &+ \text{H.c.} \nonumber \\
        &- \hbar g_0 \op{d}^\dagger\op{d} \left[\op{b}^\dagger e^{i\Omega t} + \op{b} e^{- i\Omega t} \right] & &
        \label{eq:app_H_all_extra_terms}\\
        \equiv & \;\op{H}_\text{0} + \op{H}_\text{rot} + \op{H}_\text{d} + \op{H}_\text{rf,G} + \op{H}_\text{q},& &\nonumber
    \end{alignat}
    where here and in the following, we neglect time dependencies of the amplitudes and the Hamiltonians in the notation for convenience. $\op{H}_0$ is the main interaction term and $\op{H}_\text{rot}$ is the interaction term that vanishes in the RWA. The two extra driving terms are given by $\op{H}_\text{d}$, the derivative term, and $\op{H}_\text{rf,G}$, the radiation force term induced by $G_\pm$. $\op{H}_q$ is an interaction term quadratic in the cavity quadratures. In the literature \cite{AspelmeyerRMP14, Clerk20}, one usually neglects the driving terms and also performs a so-called linearization using that in almost all cases $G_\pm \gg g_0$ to neglect the quadratic term. However, at this point we do not assume anything about the different amplitudes and keep all terms.

    For the system parameters we use, the variance of the photon number $\op{d}^\dagger\op{d}$ is small and the corresponding term in $\op{H}_\text{q}$ can be replaced by its expectation value $\braket{\op{d}^\dagger\op{d}}$, which we treat in the following as another time-dependent parameter.
    With that, we first notice that $\op{H}_\text{0}$ and $\op{H}_\text{rot}$ only involve terms in which a single annihilation or creation operator from each system appears, while the other three parts of the Hamiltonian involve only a single operator of the cavity or resonator. This has the consequence that the latter have no influence on the variance of the quadratures and thus also not on the squeezing, which means that they can be neglected when only considering squeezing. Although our numerical simulations show that this is indeed the case, we also present a mathematical proof in the following. This is done by going from the interaction picture to the Heisenberg picture and by analyzing how the different terms influence the time evolution of the quadratures.

    We start by considering the master equation for observables in the Heisenberg picture, the so-called adjoint quantum master equation \cite{Breuer02}. Note that we denote the Heisenberg picture version of observables in the following with a small letter and an explicit time dependence. For an arbitrary observable $\op{A}$, the adjoint master equation for the Heisenberg picture version $\op{a}(t)$ reads (setting $\hbar\equiv1$ in the following):
    \begin{align}
        &\deriv[\op{a}(t)]{t} = i \left[\op{H}, \op{A}\right]_H\nonumber\\
        &\quad + \sum_k \left(\op{L}_k^\dagger \op{A} \op{L}_k - \frac{1}{2}\left\{\op{L}_k^\dagger \op{L}_k, \op{A} \right\}\right)_H + \left(\del[\op{A}]{t}\right)_H,
        \label{eq:app_Lindblad_master_Heisenberg}
    \end{align}
    where the index $H$ denotes that all operators coming out of the calculation are replaced by their Heisenberg picture version at time $t$. Since the quadratures are all not explicitly time-dependent, the last term vanishes in all cases in the following. With this equation, one can start to calculate the quadratures in the Heisenberg picture.

    We define the quadrature operators (in the interaction picture) in the following for the resonator as
    \begin{equation}
        \op{X}_1 = \frac{1}{\sqrt{2}}\left(\opdag{b} + \op{b}\right);  \qquad 
        \op{X}_2 = \frac{i}{\sqrt{2}}\left(\opdag{b} - \op{b}\right),
        \label{eq:app_def_quadratures_res}
    \end{equation}
    and for the cavity as 
    \begin{equation}
        \op{Y}_1 = \frac{1}{\sqrt{2}}\left(\opdag{d} + \op{d}\right);  \qquad 
        \op{Y}_2 = \frac{i}{\sqrt{2}}\left(\opdag{d} - \op{d}\right).
        \label{eq:app_def_quadratures_cav}
    \end{equation}
    with the commutators
    \begin{equation}
        \left[\op{X}_1, \op{X}_2\right] = i = \left[\op{Y}_1, \op{Y}_2\right].\label{eq:app_commutators_quadratures}
    \end{equation}

    In the following, we apply \cref{eq:app_Lindblad_master_Heisenberg} to the quadrature operators. Therefore, it is useful to rewrite the Hamiltonian in terms of the latter. Introducing the abbreviations
    \begin{align*}
        \mathcal{G} &= \frac{\sqrt{2}}{g_0}\left[G_+^2 + G_-^2 + 2 G_+G_- \cos{2\Omega t}\right] + \sqrt{2}g_0 \expval{\opdag{d}\op{d}}\\
        \Delta G &= G_- - G_+ \\
        \overline{G} &= G_- + G_+ \\
        \Delta \alpha &= - \frac{\sqrt{2}}{g_0} \deriv[\Delta G]{t} \\
        \overline{\alpha} &= - \frac{\sqrt{2}}{g_0} \deriv[\overline{G}]{t}
    \end{align*}
    one can rewrite the Hamiltonian as
    \begin{align}
        \op{H}_{XY} =
        & -\left(\overline{G} \op{Y}_1 \op{X}_1 + \Delta G \op{Y}_2 \op{X}_2 \right)\nonumber\\
        & -\left(\overline{G} \op{Y}_1 \op{X}_1 - \Delta G \op{Y}_2 \op{X}_2 \right)\cos(2\Omega t)\nonumber\\
        &\hspace{1cm} - \left(\overline{G} \op{Y}_1\op{X}_2 + \Delta G \op{Y}_2 \op{X}_1 \right)\sin(2\Omega t)\nonumber\\
        & +\overline{\alpha} \op{Y}_2 \cos(\Omega t) - \Delta \alpha \op{Y}_1 \sin(\Omega t)\nonumber\\
        & -\mathcal{G}\op{X}_1 \cos(\Omega t) - \mathcal{G}\op{X}_2 \sin(\Omega t)\label{eq:app_H_in_quadratures}\\
        \equiv & \;\op{H}_\text{0} + \op{H}_\text{rot} + \op{H}_\text{d} + \op{H}_\text{rf}.&\nonumber
    \end{align}
    where we combined $\op{H}_\text{q}$ and $\op{H}_\text{rf,G}$ into $\op{H}_\text{rf}$ as they both act as an effective radiation pressure force on the resonator. Using the commutator relations \eqref{eq:app_commutators_quadratures}, one can calculate the commutators of the quadratures with the individual terms in the Hamiltonian. We summarized them in table \ref{tab:app_commutators}.
    One can see that the interaction terms $\op{H}_\text{0}$ and $\op{H}_\text{rot}$ only contain two quadratures, such that the commutators yield only terms with a single quadrature.
    The extra driving terms $\op{H}_\text{d}$ and $\op{H}_\text{rf}$ only involve a single quadrature operator in each term, and thus the commutator with any quadrature yields in this case a scalar multiple of the identity, which one can see in the two columns on the right.
    This implies that for all times, $\op{x}_1(t)$ only has contributions that involve a single quadrature or that are proportional to the identity.
    Further, since the identity operator is always mapped to itself, the driving terms do not contribute to the terms involving quadratures and thus have no influence on the squeezing.
    In the following, we exemplify this in a general way for the $\op{X}_1$-quadrature, since this is the one to be squeezed in the main text although the same can be done for the other three quadratures.

    For the full time evolution, one also needs to calculate the non-unitary part. We again summarized the contributions in table \ref{tab:app_non-unitary_evolution}.
    Using tables \ref{tab:app_commutators} and \ref{tab:app_non-unitary_evolution} and inserting the expressions into Eq.~\eqref{eq:app_Lindblad_master_Heisenberg}, the equation of motion for the Heisenberg picture quadrature $\op{x}_1(t)$ reads
    \begin{align}
        \deriv[\op{x}_1(t)]{t}
        = &- \op{y}_1(t) \overline{G} \sin(2\Omega t) - \op{y}_2(t) \Delta G (1 - \cos(2\Omega t))\nonumber\\
        & - \frac{\Gamma}{2}\op{x}_1(t) - \mathcal{G} \sin(\Omega t).
        \label{eq:app_x1_eq_of_motion}
    \end{align}

\begin{widetext}
\begin{center}
    \begin{table}[ht]
        \centering
        \caption{Commutators of the individual parts of the Hamiltonian with the quadrature operators of resonator and cavity}
        $\begin{array}{c||c|c|c|c}
            \op{A} & i\left[\op{H}_0, \op{A}\right] & i\left[\op{H}_\text{rot}, \op{A}\right] & i\left[\op{H}_\text{d}, \op{A}\right] & i\left[\op{H}_\text{rf}, \op{A}\right] \\\hline\hline
            \op{X}_1 & -\Delta G \op{Y}_2   & - \overline{G} \op{Y}_1 \sin(2\Omega t) + \Delta G \op{Y}_2 \cos(2\Omega t) & 0 & - \mathcal{G} \sin(\Omega t)     \\
            \op{X}_2 & \overline{G}\op{Y}_1 & + \overline{G} \op{Y}_1 \cos(2\Omega t) + \Delta G \op{Y}_2 \sin(2\Omega t) & 0 &  \mathcal{G} \cos(\Omega t)      \\
            \op{Y}_1 & -\Delta G \op{X}_2   & - \Delta G \op{X}_1 \sin(2\Omega t)     + \Delta G \op{X}_2 \cos(2\Omega t) & \overline{\alpha} \cos(\Omega t) & 0 \\
            \op{Y}_2 & \overline{G}\op{X}_1 & + \overline{G} \op{X}_1 \cos(2\Omega t) + \overline{G} \op{X}_2 \sin(2\Omega t) & \Delta \alpha \sin(\Omega t) & 0
        \end{array}$
        \label{tab:app_commutators}
    \end{table}
\end{center}
\end{widetext}

    Similar expressions can be obtained for $\op{x}_2(t)$, $\op{y}_1(t)$ and $\op{y}_2(t)$. Now one can use that at $t=0$, i.e., before switching on the driving fields, all operators coincide in Heisenberg and Schrödinger picture. This allows to divide the time evolution into two parts, a part which is a linear combination of the four quadratures with time-dependent weights and one part proportional to the identity:
    \begin{equation}
        \op{x}_1(t) = \op{\chi}_1(t) + R_1(t)\label{eq:app_x1}
    \end{equation}
    with 
    \begin{equation*}
        \op{\chi}_1(t) = (1 + a_1(t))\op{X}_1 + a_2(t)\op{X}_2 + b_1(t)\op{Y}_1 + b_2(t)\op{Y}_2.
    \end{equation*}
    where $a_1(t), a_2(t), b_1(t), b_2(t)$ and $R_1(t)$ are some real functions of time with initial value zero. Please note that we put the time dependency into the scalar functions and used the four time-independent quadratures (and the identity) as a basis.

    Assuming a structure for the other quadratures similar to \cref{eq:app_x1},
    \begin{align*}
        \op{x}_2(t) &= \op{\chi}_2(t) + R_2(t)\\
        \op{y}_1(t) &= \op{\gamma}_1(t) + C_1(t)\\
        \op{y}_2(t) &= \op{\gamma}_2(t) + C_2(t),
    \end{align*}
    and inserting this into Eq. \eqref{eq:app_x1_eq_of_motion}, one can observe that the differential equation separates into two independent ones:
    \begin{align}
        \deriv[\op{\chi}_1(t)]{t}
        =& - \op{\gamma}_1(t) \overline{G} \sin(2\Omega t) - \op{\gamma}_2(t) \Delta G (1 - \cos(2\Omega t)) \nonumber\\
        &- \frac{\Gamma}{2}\op{\chi}_1(t),\\
        \deriv[R_1(t)]{t}
        =& - C_1(t) \overline{G} \sin(2\Omega t) - C_2(t) \Delta G (1 - \cos(2\Omega t)) \nonumber\\
        &- \frac{\Gamma}{2}R_1(t) - \mathcal{G} \sin(\Omega t).\label{eq:app_deriv_R1}
    \end{align}
    Now it becomes apparent that the extra driving terms $\op{H}_\text{d}$ and $\op{H}_\text{rf}$ indeed only influence $R_1(t)$ resp. $R_2(t), C_1(t), C_2(t)$, \ie only the terms proportional to the identity. 
    
    In the variance, the square of the expectation value appears. It reads
    \begin{equation}
        \Expval{\op{x}_1(t)}^2 = \Expval{\op{\chi}_1(t)}^2 + 2 R_1(t)\Expval{\op{\chi}_1(t)} + R_1^2(t).\label{eq:app_expval_x1}
    \end{equation}

    \begin{table}[t]
        \centering
        \caption{Non-unitary part of the evolution of the quadratures and the squared quadratures. Here, we defined the dissipative part as $\supop{L}\op{A}=i[\op{H},\op{A}] + \supop{L}_d\op{A}$}
        $\begin{array}{c|c||c|c}
            \op{A} & \supop{L}_d\op{A} & \op{A}^2 & \supop{L}_d\op{A}^2\\\hline\hline
            \op{X}_1   & -\Gamma \op{X}_1/2 & \op{X}_1^2 & -\Gamma (\op{X}_1^2 - (n_\mathrm{th}+1/2)) \\
            \op{X}_2   & -\Gamma \op{X}_2/2 & \op{X}_2^2 & -\Gamma (\op{X}_2^2 - (n_\mathrm{th}+1/2)) \\
            \op{Y}_1   & -\kappa \op{Y}_1/2 & \op{Y}_1^2 & -\kappa (\op{Y}_1^2 - 1/2 ) \\
            \op{Y}_2   & -\kappa \op{Y}_2/2 & \op{Y}_2^2 & -\kappa (\op{Y}_2^2 - 1/2 )
        \end{array}$
        \label{tab:app_non-unitary_evolution}
    \end{table}
    
    We now consider the second part of the variance, i.e., the squared operator. The equation of motion reads
    \begin{align}
        \deriv[\op{x}_1^2(t)]{t} 
        =& \; 2 \left(\op{X}_1 i\left[\op{H}(t), \op{X}_1\right]\right)_H\nonumber\\
        &- \Gamma \left(\op{x}_1^2(t) - (n_\mathrm{th} + 1/2)\right)\nonumber\\
        =& \; 2 \left(\op{X}_1 \left\{ i\left[\op{H}(t), \op{X}_1\right] -\frac{\Gamma}{2}\op{X}_1 \right\}\right)_H\nonumber\\
        &+ \Gamma(n_\mathrm{th} + 1/2).\label{eq:app_Heisenberg_x1sq}
    \end{align}
    This equation looks very similar to 
    \begin{equation*}
        \deriv[\op{x}_1(t)^2]{t} = 2 \op{x}_1(t) \deriv[\op{x}_1(t)]{t}.
    \end{equation*}
    However, an extra constant arises from the decay on the right-hand side of \cref{eq:app_Heisenberg_x1sq}. This demonstrates that in general the squared operators evolve differently than the squares of the evolved operators for systems undergoing dissipation. Therefore, we consider the constant in the following as an inhomogeneity in the differential equation and show that the homogeneous equation is indeed solved by $\op{x}_1^2(t) = (\op{x}_1(t))^2$. To see this, we write out the commutators. With that, \cref{eq:app_Heisenberg_x1sq} becomes
    \begin{align}
        \deriv[\op{x}_1^2(t)]{t} =
        & - 2 [\op{x}_1\op{y}_{2}](t) \Delta G (1 - \cos(2\Omega t))\nonumber\\
        & - 2 [\op{x}_1\op{y}_{1}](t) \overline{G} \sin(2\Omega t) - \Gamma \op{x}_1^2(t)\nonumber\\
        & - 2\op{x}_1(t) \mathcal{G} \sin(\Omega t) + \Gamma (n_\mathrm{th} + 1/2).\label{eq:app_Heisenberg_x1sq_explicit}
    \end{align}
    where $[\op{x}_1\op{y}_{1/2}](t) = \left(\op{X}_1 \op{Y}_{1/2}\right)_H$. Note that $\op{x}_1(t)$ is known from \cref{eq:app_x1_eq_of_motion} and independent of the other variables in \cref{eq:app_Heisenberg_x1sq_explicit}. Now we make the ansatz
    \begin{equation}
        \op{x}_1^2(t) = (\op{x}_1(t))^2 + \gamma_{11}(t)\label{eq:app_ansatz_x1sq}
    \end{equation}
    and similar for the other products of operators
    \begin{align}
        [\op{x}_i\op{x}_j](t) &= \op{x}_i(t)\op{x}_j(t) + \gamma_{ij}(t)\label{eq:app_ansatz_xi_xj}\\
        [\op{x}_i\op{y}_j](t) &= \op{x}_i(t)\op{y}_j(t) + \beta_{ij}(t)\label{eq:app_ansatz_xi_yj}\\
        [\op{y}_i\op{y}_j](t) &= \op{x}_i(t)\op{x}_j(t) + k_{ij}(t)\label{eq:app_ansatz_yi_yj}
    \end{align}
    where $i,j \in \{1,2\}$ and $\gamma_{ij}(t), \beta_{ij}(t), k_{ij}(t)$ are real functions proportional to the identity with an initial value of zero. What is left to show is that these actually fulfill the differential equations. We again take $\op{x}_1^2(t)$ as an example. Inserting \cref{eq:app_ansatz_xi_xj,eq:app_ansatz_xi_yj,eq:app_ansatz_yi_yj} into \cref{eq:app_Heisenberg_x1sq_explicit} yields:
    \begin{align}
        \deriv[\op{x}_1^2(t)]{t}
        =& \; 2 \op{x}_1(t) \Bigl(- \op{y}_{2}(t) \Delta G (1 - \cos(2\Omega t)) - \frac{\Gamma}{2} \op{x}_1(t) \nonumber\\
        &\hspace{2cm} - \op{y}_{1}(t) \overline{G} \sin(2\Omega t) - \mathcal{G} \sin(\Omega t)\Bigr)\nonumber\\
        & + \Bigl(- 2\beta_{11}(t) - 2\beta_{12}(t) - \Gamma\gamma_{11}(t)\nonumber\\
        &\hspace{3.2cm} + \Gamma (n_\mathrm{th} + 1/2)\Bigr).\label{eq:app_Heisenberg_x1sq_w_ansatz}\\
        \overset{!}{=}& \; 2 \op{x}_1(t) \deriv[\op{x}_1(t)]{t} + \deriv[\gamma_{11}(t)]{t}.
    \end{align}
    The bracket in the first two lines of \cref{eq:app_Heisenberg_x1sq_w_ansatz} exactly coincides with \cref{eq:app_x1_eq_of_motion} which justifies the ansätze in \cref{eq:app_ansatz_xi_xj,eq:app_ansatz_xi_yj,eq:app_ansatz_yi_yj}. Further, this yields a differential equation for $\gamma_{11}(t)$ as
    \begin{equation}
        \deriv[\gamma_{11}(t)]{t} = - 2\beta_{11}(t) - 2\beta_{12}(t) - \Gamma\gamma_{11}(t) + \Gamma (n_\mathrm{th} + 1/2).\label{eq:app_deriv_gamma11}
    \end{equation}
    Similar equations can be written down for the other products of two quadratures. This yields an important feature of the solutions $\gamma_{ij}(t)$, $\beta_{ij}(t)$ and $k_{ij}(t)$. Since they are all initially zero and obey a differential equation similar to \cref{eq:app_deriv_gamma11}, they are indeed only proportional to the identity and do not involve any quadratures. More importantly, it also shows that $\gamma_{11}(t)$ is independent of the driving terms.

    Finally, we can calculate the variance of the $\op{X}_1$-quadrature in the Heisenberg picture. Using \cref{eq:app_expval_x1} and \cref{eq:app_ansatz_x1sq} yields
    \begin{align}
        \Delta X_1^2(t) 
        =& \expval{\op{x}_1^2(t)} - \expval{\op{x}_1(t)}^2\nonumber\\\
        =& \Expval{\op{\chi}_1(t)^2 + 2 R_1(t)\op{\chi}_1(t) + R_1^2(t) + \gamma_{11}(t)}\nonumber\\
         & - \left(\Expval{\op{\chi}_1(t)}^2 + 2 R_1(t)\Expval{\op{\chi}_1(t)} + R_1^2(t)\right)\nonumber\\
        =& \Expval{\op{\chi}_1(t)^2} - \Expval{\op{\chi}_1(t)}^2 + \gamma_{11}(t).
    \end{align}
    As shown earlier, $\op{\chi}_1(t)$ and $\gamma_{11}(t)$ are independent of the driving terms. This in turn means that the three driving parts $\op{H}_\text{d}$, $\op{H}_\text{rf,G}$ and $\op{H}_\text{q}$ in \cref{eq:app_H_all_extra_terms} have no influence on the squeezing and can thus be neglected. This yields the final Hamiltonian in \cref{eq:H_final_with_rotating_terms}.


\bibliography{references}

\end{document}